\documentclass[showpacs]{revtex4}
\usepackage{epsfig}
\def\beq{\begin{equation}}
\def\eeq{\end{equation}}
\def\beqa{\begin{eqnarray}}
\def\eeqa{\end{eqnarray}}

\def\GeV{\nobreak\,\mbox{GeV}}

\def\pli{p^\prime}

\def\mli{{M^\prime}^2}

\begin{document}
\title{$D^{*}D \rho $ vertex from QCD sum rules}
\author{B. Os\'orio Rodrigues}
\affiliation{Instituto de F\'{\i}sica, Universidade do Estado do Rio de 
Janeiro, Rua S\~ao Francisco Xavier 524, 20550-900 Rio de Janeiro, RJ, Brazil}
\author{M.E.  Bracco}
\affiliation{Faculdade de Tecnologia, Universidade do Estado do Rio de 
Janeiro, Rod. Presidente Dutra Km 298, Polo Industrial,     , Resende, RJ, Brazil}
\author{ M. Nielsen and F.S. Navarra}
\affiliation{Instituto de F\'{\i}sica, Universidade de S\~{a}o Paulo, 
C.P. 66318, 05389-970 S\~{a}o Paulo, SP, Brazil.}

\begin{abstract}
We calculate the  form factors  and the coupling constant in  the $D^{*}D \rho $ vertex 
in the framework of QCD sum rules. We evaluate  the 
three point correlation functions of the vertex considering both $ D $ and $ \rho $ 
mesons off--shell. The form factors obtained are very different but  give 
the same coupling constant: $g_{D^{*}D \rho } = 4.1  \pm 0.1$ GeV$^{-1}$.
\end{abstract}


\maketitle

\section{Introduction}

Over the last years the strong interaction of charmed hadrons among themselves and with 
other species of hadrons has received increasing attention. From the discovery of charmed 
mesons in the seventies until the late eighties there was no motivation to study in 
detail the interactions of these particles. In the early nineties there was a series of 
papers trying 
to compute the cross section of a $J/\psi$ with ordinary light hadrons. The motivation 
came from the heavy ion program running at CERN and later at RHIC. At that time $J/\psi$ 
suppression was considered as a signature of quark gluon plasma (QGP) formation and it was 
very important to know as accurately as possible the purely hadronic (non-QGP induced) 
charmonium suppression, which would be the background for the QGP signal. From 2000 on, it
became slowly clear that the physics of $J/\psi$ is much more complex than thought before and 
its simple suppression was no longer considered as a QGP signal and the subject lost 
interest. On the other hand, during those years, at the  $B$ factories the collaborations 
BABAR and BELLE started to produce results. One of the important decay channels  of the $B$ 
mesons is into $J/\psi$ (plus other things).  Moreover these collaborations found new 
charmonium states (the $X$, the $Y$'s and the $Z$), which also decay into $J/\psi$ or 
into $\psi'$. It has 
been conjectured that both $B$ and the new charmonium  states very often decay into an 
intermediate two body state with $D$'s and/or $D^*$'s, which then undergoes final state 
interactions, with the exchange of one or more virtual mesons. In order to calculate the 
amplitudes of these processes we need to know the relevant vertices involving the charmed
mesons. As an example of  specific situation where a precise knowledge of the 
$  D^*  D  \rho  $ form factor is required, we may consider the decay 
$X(3872) \, \rightarrow \, J/\psi \, + \, \rho$. As suggested in \cite{lzz}, this decay 
proceeds in two steps. First the $X$ decays into a $D$ - $D^*$ intermediate state and then
these two particles exchange a $D^*$ producing the final $J/\psi$ and $\rho$. This is 
shown  in Fig. 1b and 1f of   \cite{lzz}. In order to compute the effect of these 
interactions in the final decay rate we need the $  D^*  D   \rho  $ form factor. More 
generally, we need to know all the charm form factors to correclty calculate the 
interaction of $J/\psi$ with light hadrons and the final state interactions in $B$ decays. 
These form factors have been calculated in the framework of QCD sum rules (QCDSR) 
\cite{svz} techniques in a series of works on vertices involving charmed mesons, namely 
$D^* D \pi$ \cite{nnbcs00,nnb02}, 
$D D \rho$\cite{bclnn01},  $D D J/\psi$ \cite{mnns02},  
$D^* D J/\psi$ \cite{mnns05},  
$D^* D^* \pi$ \cite{wang,cdnn05}, $D^* D^* J/\psi$ \cite{bcnn05},   
$D_s D^* K$, $D_s^* D K$ \cite{bccln06} ,  $D D \omega$  \cite{hmm07} and 
$D^* D^* \rho$ \cite{bcnn08}.

In the present paper we calculate the $D^{*}D \rho$ form factor with QCDSR.   
In the next section, for completeness we describe the QCDSR technique and in 
section III we present the  results and compare them with results obtained in 
other works.  

\section{The sum rule for the  $ D^{*}D \rho $  vertex}

Following our previous works and especially Ref. \cite{mnns05}, 
we write the three-point function associated with the $D^{*}D \rho $ vertex, 
which is given by
\begin{equation}
\Gamma_{\mu \nu}^{(D)}(p,\pli)=\int d^4x \, d^4y \;\;
e^{i\pli\cdot x} \, e^{i(\pli-p)\cdot y}
\langle 0|T\{j_{\mu}^{{{\rho}}}(x) j^{D }(y) 
 j_{\nu}^{{D^{*} \dagger}}(0)|0\rangle\, 
\label{corredoff} 
\end{equation}
for an off-shell $D$ meson, and:
\begin{equation}
\Gamma_{\mu \nu}^{({\rho})}(p,\pli)=\int d^4x \, 
d^4y \;\; e^{i\pli\cdot x} \, e^{i(\pli-p)\cdot y}\;
\langle 0|T\{j^{D }(x)  j_{\mu}^{{\rho} }(y) 
 j_{\nu}^{D^{* \dagger}} (0)\}|0\rangle\, 
\label{correrhooff} 
\end{equation}
for an off-shell ${\rho}$ meson. The general expression for the vertices 
(\ref{corredoff}) and (\ref{correrhooff}) has only one Lorentz structure. 
Equations~(\ref{corredoff}) and (\ref{correrhooff}) can be calculated in two 
diferent ways: using quark 
degrees of freedom --the theoretical or QCD side-- or using hadronic 
degrees of freedom --the phenomenological side. In the QCD side the 
correlators are evaluated  using the 
Wilson operator product expansion (OPE). The OPE incorporates the effects 
of the QCD vacuum through an infinite series of condensates of in\-crea\-sing 
dimension. On the other hand, the representation in terms of 
hadronic degrees of freedom is responsible for the introduction of the form 
factors, decay constants and masses. Both representations are matched invoking 
the quark-hadron global duality.

\subsection{The phenomenological side}

The $ D^{*}D \rho $ vertex can be studied with hadronic degress of freedom. The 
corresponding three-point functions,  Eqs.~(\ref{corredoff}) and 
(\ref{correrhooff}), are written in terms of hadron masses, decay constants and
form factors. This is the so called phenomenological side of the sum rule and it 
is based on the interactions at the hadronic level, which  are described here by the   
following effective Lagrangian  \cite{linko,su}
\beq
\mathcal{L}_{D^{*}D \rho }=
 - \, g_{D^{*}D \rho } \,\, \epsilon^{\gamma \delta\alpha\beta} \,  
\Big ( \, {D} \, \partial_{\gamma}  \rho_{\delta} \,  \partial_{\alpha}  \bar{D^*_{\beta}} 
                  + hc \Big ) 
\label{lagran}
\eeq
from where one can extract the matrix element associated with the
$D^{*}D \rho $ vertex. In the above expression we have $\epsilon^{0123}=+1$. 
Saturating Eqs.~(\ref{corredoff}) and (\ref{correrhooff}) with the appropriate 
$D $, $D^*$ and  $\rho$ states and making all the contractions we arrive at: 
\beq
\Gamma^{ (M)}_{\mu\nu}(p,\pli)=\Lambda^{(M)}_{phen}(p^2,{\pli}^2,q^2) \, 
\epsilon_{\alpha\beta \mu\nu}p^\alpha{\pli}^\beta  + \mbox{h.~r.}
\label{phen}
\eeq
where h.~r. means higher resonances and $q = p - \pli$. 
The invariant amplitude $\Lambda^{(M)}$ is given by
\beq
\Lambda_{phen}^{(D )}= -g^{(D)}_{D^*D \rho }(Q^2) 
  \frac{ f_{D} f_{\rho} f_{D^*} \frac{m^2_D}{m_c} m_{D^*} m_{\rho} }
{(P^2+m^2_{D^*}) (Q^2+m^2_{D}) ({P^\prime}^2 +m^2_{\rho}) } 
\label{phendoff}
\eeq
for an off-shell $D$ meson and 
\beq
\Lambda_{phen}^{(\rho)}= -g^{(\rho)}_{D^{*}D \rho }(Q^2) 
\frac{f_{D} f_{\rho} f_{D^*} \frac{m^2_{D}}{m_c} m_{D^*} m_{\rho} }
{(P^2+m^2_{D^*})(Q^2+m^2_{\rho})({P^\prime}^2 +m^2_{D})} 
\label{phenrhooff}
\eeq
for an off-shell $\rho$ meson. In the above expressions $P^2 = - p^2$, 
${P^\prime}^2 = - {p^\prime}^2$ and $Q^2 = -q^2$.

The meson decay constants appearing in the equations above  
are defined by the vacuum to meson transition amplitudes:
\beq
\langle 0|j^{D}|D\rangle={m_D^2f_{D}\over m_c}\;,
\label{fd}
\eeq
and
\beq
\langle V(p,\epsilon)|j^\dagger_\alpha|0\rangle=m_{V}f_{V}\epsilon^*_\alpha
\; ,
\label{fv}
\eeq
for the vector mesons $V=D^{*}$ and $V=\rho$. The form factor which we want
to estimate is defined through the vertex function for an off-shell $\rho$ meson:
\beq
\langle D^{*}(p,\lambda)|D(\pli)\rho(q,\lambda^\prime)\rangle
= i \, g_{D^{*} D \rho}^{(\rho)}(q^2) \, 
\epsilon^{\alpha\beta\gamma\delta}\epsilon_\alpha^\lambda(p)
\epsilon_\gamma^{\lambda^\prime}(q)\pli_{\beta}q_{\delta}\;.
\label{ver}
\eeq
where $\epsilon_\alpha^\lambda(p)$ and $\epsilon_\gamma^{\lambda^\prime}(q)$ are the 
polarization vectors associated with the $D^{*}$ and $\rho$ respectively. 
An analogous expression holds for an off-shell $D$ meson. As it will be seen in the next
subsection, the contribution of higher resonances and continuum in 
Eq.~(\ref{phen}) will be transferred to the OPE side.

\subsection{The  OPE  side}

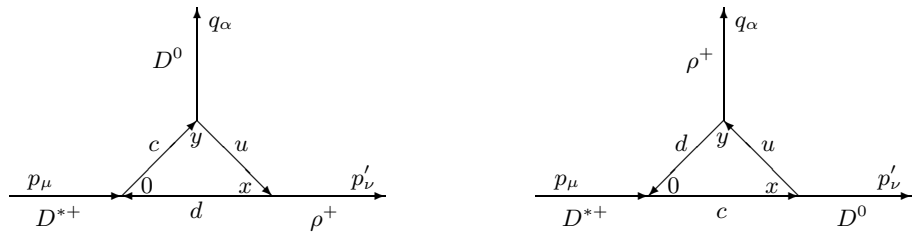
\begin{figure}[h]
\begin{picture}(12,3.5)
\put(0.0,0.5){\vector(1,0){1.5}}
\put(3.5,0.5){\vector(-1,0){2}}
\put(3.5,0.5){\vector(1,0){1.5}}
\put(1.5,0.5){\vector(1,1){1}}
\put(2.5,1.5){\vector(1,-1){1}}
\put(2.5,1.5){\vector(0,1){1.5}}
\put(2.65,2.75){$q_\alpha$}
\put(0.25,0.65){$p_\mu$}
\put(4.55,0.65){$p'_\nu$}
\put(2.4,0.2){$ d$}
\put(1.85,1.1){$c$}
\put(3,1.1){$u$}
\put(2.4,1.2){$y$}
\put(1.75,0.53){$0$}
\put(3.05,0.53){$x$}
\put(1.9,2.2){$D^0$}
\put(0.35,0.1){${D^{*+}}$}
\put(4,0.1){${\rho}^+$}
\put(7,0.5){\vector(1,0){1.5}}
\put(8.5,0.5){\vector(1,0){2}}
\put(10.5,0.5){\vector(1,0){1.5}}
\put(9.5,1.5){\vector(-1,-1){1}}
\put(10.5,0.5){\vector(-1,1){1}}
\put(9.5,1.5){\vector(0,1){1.5}}
\put(9.65,2.75){$q_\alpha$}
\put(7.25,0.65){$p_\mu$}
\put(11.55,0.65){$p'_\nu$}
\put(9.4,0.2){$ c$}
\put(8.85,1.1){$ d$}
\put(10,1.1){$ u$}
\put(9.4,1.2){$y$}
\put(8.75,0.53){$0$}
\put(10.05,0.53){$x$}
\put(9.00,2.2){${\rho^+}$}
\put(7.35,0.1){${D^{*+}}$}
\put(11,0.1){${D^0}$}
\end{picture}
\caption{Perturbative diagrams for the $D$ off-shell (left) and $\rho$
off-shell (right) correlators.}
\label{fig1}
\end{figure}

In the OPE or theoretical side  each meson interpolating field appearing in 
Eqs.~(\ref{corredoff}) and (\ref{correrhooff}) is written 
in terms of the quark field operators in the following form: 
\beq
j_{\mu}^{\rho}(x) = \bar d(x) \gamma_{\mu} u(x);
\label{corho}
\eeq 
\beq
j^{D}(y) = i \bar u(y) \gamma_{5} c(y) 
\label{coD}
\eeq
and 
\beq
j_{\nu}^{D^*}(0) = \bar d(0) \gamma_{\nu} c(0) 
\label{coDs}
\eeq
where $u$, $d$  and $c$ are the up, down and charm quark field respectively. 
Each  one of these currents has the same quantum numbers of the associated
meson. The correlators (\ref{corredoff}) and (\ref{correrhooff}) 
receive contributions from all terms in the OPE. The first (and dominant) of these 
contributions comes from the perturbative term and it is represented in Fig.~\ref{fig1}.
Here we will consider the perturbative diagram and the quark condensate. We can write  
$\Gamma_{\mu\nu}$ in terms of the invariant amplitude:
\beq
\Gamma^{(M)}_{\mu\nu}(p,\pli)=\Lambda^{(M)}_{OPE} (p^2,{\pli}^2,q^2) \, 
\epsilon_{\alpha\beta
\mu\nu}p^\alpha{\pli}^\beta\;,
\eeq
where the meson $M (= D  \, , \, \rho)$ is off-shell.  
We can write a double dispersion relation for $\Lambda$,
over the virtualities $p^2$ and ${\pli}^2$ holding $q^2$  fixed:
\beq
\Lambda^{(M)}_{OPE}(p^2,{\pli}^2,q^2)=-{1\over4\pi^2} \int  ds
\int  du~ {\rho^{(M)}(s,u,t)\over(s-p^2)(u-{\pli}^2)} \, + \, 
\Lambda^{(M)}_{\langle \bar q q\rangle }
\label{dis}
\eeq
where $t = q^2 $ and  $\rho^{(M)}(s,u,t)$ is the double discontinuity of the amplitude
$\Lambda^{(M)}(p^2,{\pli}^2,q^2)$ when the meson $M (= D  \, , \, \rho)$ is off-shell.  
The perturbative contribution to the double discontinuity in (\ref{dis}) 
for an off-shell $D$ meson is given by:
\beq
\rho^{(D)}(u,s,t)=
\frac{3m_c}{\sqrt{\lambda}}
\left[\frac{u (2 m_c^2 -s -t +u)}{\lambda}\right]
\label{rhodoff}
\eeq
with $\lambda =(u+s-t)^2 -4us$. The integration limits in the integrals in (\ref{dis}) are:
$$
0 \, < \, u \, < \, \frac{m_c^2 (s + t - m_c^2) - st}{m_c^2}
$$ 
and 
$$
m_c^2 \, < \, s \, < \, s_0
$$
Evaluating the perturbative contribution for the double discontinuity for an off-shell 
$\rho$ meson we find:
\beq
\rho^{(\rho)}(u,s,t)=
\frac{3m_c \, t}{\lambda^{3/2}} \left[u + s - t - 2 m_c^2 \right]
\label{rhorhooff}
\eeq
and the corresponding integration limits in (\ref{dis}) are: 
$$ 
\frac{m_c^2 (s  - t  - m_c^2) }{s - m_c^2}  \, < \, u \, < \, u_0
$$
and 
$$
m_c^2 \, < \, s \, < \, s_0
$$
As usual, we have already transferred the 
continuum contribution from the hadronic side to the QCD side, through 
the introduction of the continuum thresholds $s_0$ and $u_0$ \cite{io2}. In doing so 
we made  the assumption that at very large values of $s$ and $u$ the double 
discontinuity appearing in the phenomenological side coincides with that of the OPE side. 
This assumption is often called quark-hadron duality.

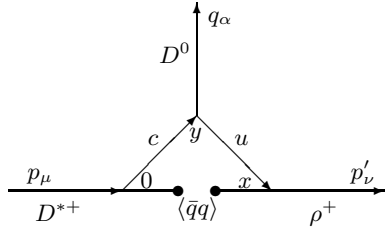
\begin{figure}[h]
\begin{picture}(6,3.5)
\put(0,0.5){\vector(1,0){1.5}}
\put(1.5,0.5){\line(1,0){0.75}}
\put(2.75,0.5){\line(1,0){0.75}}
\put(2.25,0.5){\circle*{0.15}}
\put(2.75,0.5){\circle*{0.15}}
\put(3.5,0.5){\vector(1,0){1.5}}
\put(1.5,0.5){\vector(1,1){1}}
\put(2.5,1.5){\vector(1,-1){1}}
\put(2.5,1.5){\vector(0,1){1.5}}
\put(2.65,2.75){$q_\alpha$}
\put(0.25,0.65){$p_\mu$}
\put(4.55,0.65){$p'_\nu$}
\put(2.225,0.17){$\langle \bar q q\rangle$}
\put(1.85,1.1){$c$}
\put(3,1.1){$u$}
\put(2.4,1.2){$y$}
\put(1.75,0.53){$0$}
\put(3.05,0.53){$x$}
\put(2.0,2.2){$D^0$}
\put(0.35,0.1){$D^{*+}$}
\put(4,0.1){$\rho^+$}
\end{picture}
\caption{Contribution of the $q	\bar q$ condensate to the $D$ off-shell
correlator.}
\label{fig2}
\end{figure}

In order to improve the matching between the two sides of the sum rules
we perform a double Borel transformation \cite{io2} in the variables 
$P^2=-p^2\rightarrow M^2$ and $P'^2=-{\pli}^2\rightarrow M'^2$, 
on the invariant amplitude $\Lambda_{OPE}$ and also on $\Lambda_{phen}$. Incidentally, 
this double Borel transform will kill the contribution of the quark condensate 
$\Lambda^{(\rho)}_{\langle \bar q q\rangle }$ leaving only 
$\Lambda^{(D)}_{\langle \bar q q\rangle }$, which  is represented in Fig.~\ref{fig2}  
and  is given by:
\begin{equation}
\Lambda_{\langle \bar q q\rangle}^{(D)} =  -  \langle \bar q q \rangle \,  e^{- m_c^2/M^2} 
\end{equation}
where $ \langle \bar q q\rangle $ is the light quark condensate. 

\subsection{The sum rule}

After performing the Borel transformation \cite{io2} 
on both invariant amplitudes $\Lambda^{(M)}_{OPE}$ and $\Lambda^{(M)}_{phen}$
we identify  (\ref{dis}) with (\ref{phendoff}) and then with 
(\ref{phenrhooff}). In doing so  we obtain two equations (the sum rules) for the 
form factors 
$g^{(D)}_{D^{*}D \rho}(Q^2)$  and $g^{(\rho)}_{D^{*}D \rho}(Q^2)$ respectively.  
We get the following sum rules:
\beqa
C~\frac{g_{D^{*} D \rho}^{(\rho)}(q^2)}{(q^2- m_{\rho}^2)}
e^{-\frac{m_{D}^2}{\mli}}
e^{-\frac{m_{D^{*}}^2}{M^2}}
&=&\frac{1}{4\pi^2} \int ds  \int du~
\rho^{(\rho)}(u,s,t) \, e^{-\frac{s}{M^{2}}}e^{-\frac{u}{\mli}}
\label{ffrho}
\eeqa
and 
\beqa
C~\frac{g_{D^{*} D \rho}^{(D)}(q^2)}{(q^2- m_{D}^2)}
e^{-\frac{m_{\rho}^2}{\mli}}
e^{-\frac{m_{D^{*}}^2}{M^2}}
&=&\frac{1}{4\pi^2} \int ds  \int du~
\rho^{(D)}(u,s,t) \, e^{-\frac{s}{M^{2}}}e^{-\frac{u}{\mli}} \, + \,  
\langle \bar q q \rangle \,  e^{- m_c^2/M^2} 
\label{ffd}
\eeqa
where $C={m_D^2f_D\over m_c}m_{\rho}f_{\rho}m_{D^{*}}f_{D^{*}}$.

\section{Results and discussion}

Table \ref{tableparam} shows the values of the parameters used in the present 
calculation. We used the experimental value for  $f_{\rho}$ \cite{pdg}, 
and took $f_{D}$ and $f_{D^*}$ from Ref.~\cite{wang}.
The continuum thresholds are given by 
$s_0=(m_i + \Delta_s)^2$ and $u_0=(m_o + \Delta_u)^2$, where $m_i$ and $m_o$
are the masses of the incoming and outgoing meson respectively.

\begin{table}[h]
\begin{tabular}{|c|c|c|c|c|c|c|c|} \hline
$m_c (\GeV)$ & $m_{D*} (\GeV)$ & $m_{D} (\GeV)$ & $m_{\rho} (\GeV)$ & $f_{D*} (\GeV)$ 
&$f_{D} (\GeV)$ & $f_{\rho} (\GeV)$ & 
$\langle \bar q q \rangle (\GeV)^3$  \\ \hline 
1.27& 2.01 &1.86 &0.775 & 0.240&0.170& 0.161 &$(-0.23)^3$ \\ \hline
\end{tabular}

\caption{Parameters used in the calculation.}
\label{tableparam}
\end{table}

In this work we use the following relations between the Borel masses $M^2$ and 
$M'^2$: $\frac{M^2}{M'^2} = \frac{m^2_{D^*}-m^2_c}{m^2_{\rho}}$ for a  $D$ off-shell  
and $\frac{M^2}{M'^2} = \frac{m^2_{D^*}}{m^2_{D}}$ for a $\rho$  off-shell.

\begin{figure}
\vspace{1.0cm}
\centerline{\psfig{figure=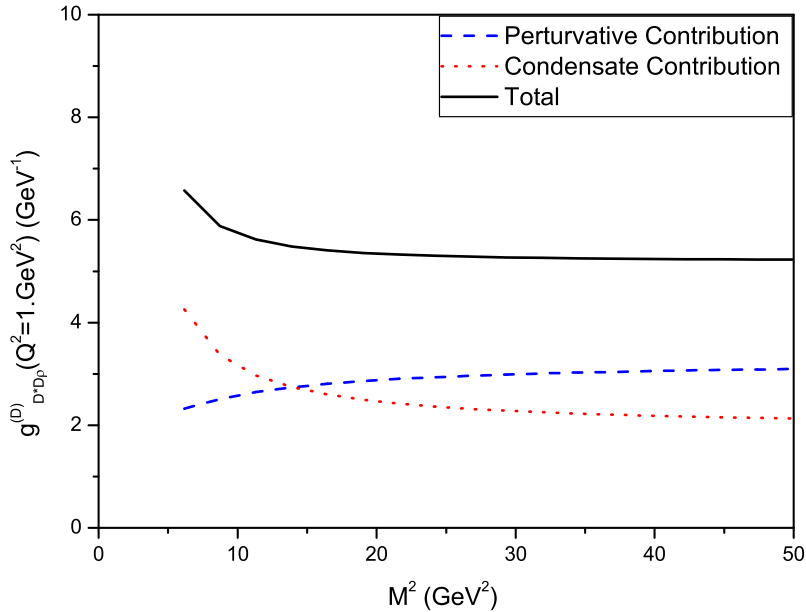,width=12cm}} 
\vspace{0.5cm}
\caption{$g^{(D)}_{D^{*}D \rho}(Q^2=1.0 \, GeV^2)$ as a function of the Borel mass $M^2$.}
\label{fig3}
\end{figure}

\begin{figure}
\vspace{1.0cm}
\centerline{\psfig{figure=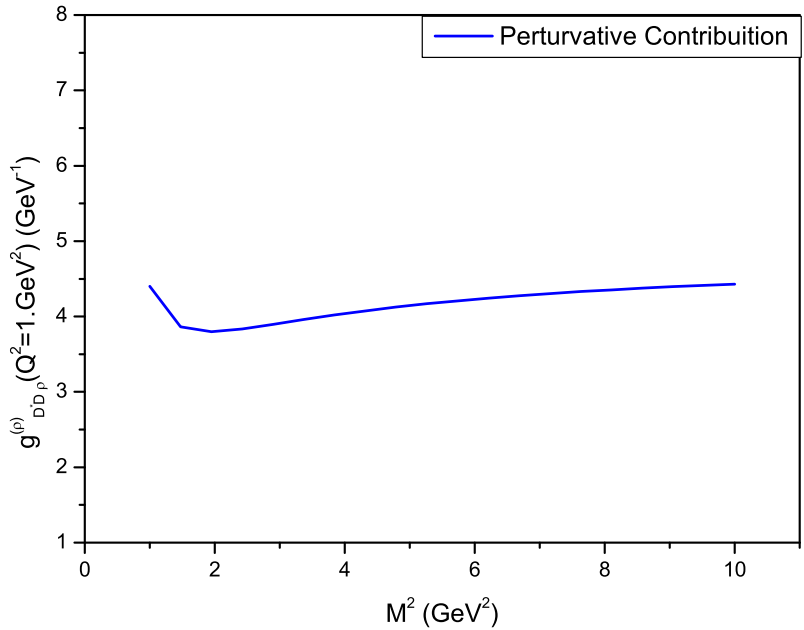,width=12cm}} 
\vspace{0.5cm}
\caption{$g^{(\rho)}_{}(Q^2=1\GeV ^2)$ as a
function of the Borel mass.}
\label{fig4}
\end{figure}

\begin{figure}[ht] 
\centerline{\psfig{figure=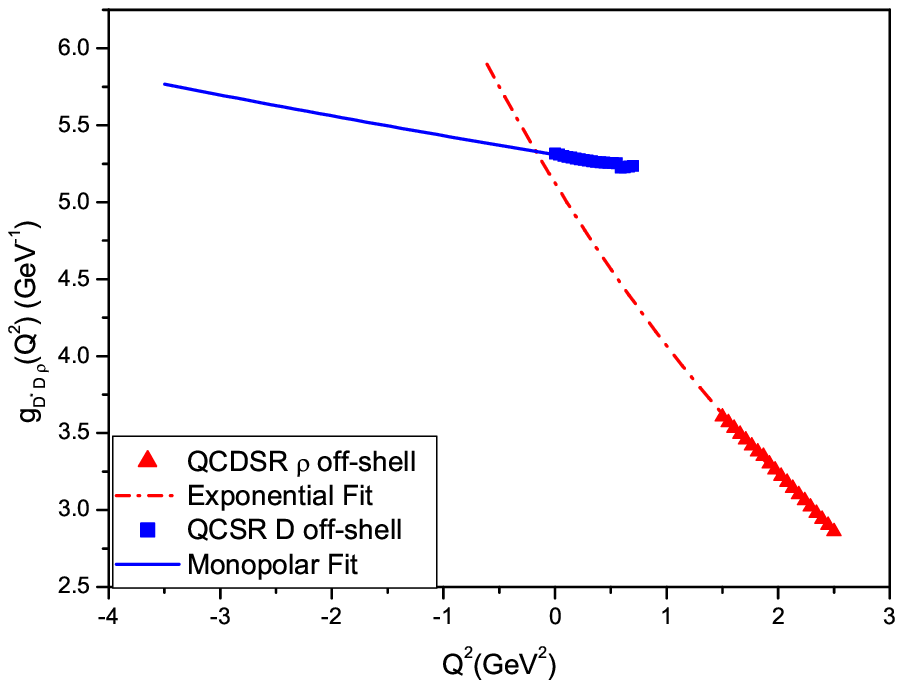,width=12cm}} 
\caption{$g^{(D)}_{D^{*}D \rho}$ (squares) and 
$g^{(\rho)}_{D^{*}D \rho}$ (triangles) QCDSR form factors as a function of
$Q^2$. The dot-dashed and solid lines 
correspond to the exponential and monopole parametrizations respectively.}
\label{fig5}
\end{figure}

Using  $\Delta_s=0.5$ GeV  and  $\Delta_u = 0.70 \GeV $ for the continuum thresholds
and fixing $Q^2=1 \GeV^2$, we found a sum rule for $g_{D^{*}D \rho}^{(D)}$ as a 
function of  $M^2$ which is very stable with respect to $M^2$ in the interval 
$ 20 < M^2 < 50 \GeV^2$. This can be seen in  Fig.~\ref{fig3}. In what follows we choose 
the value $M^2 = 30$ GeV$^2$ as a reference. In  Fig.~\ref{fig4} we show the $M^2$ 
dependence of the form factor $g_{D^{*}D \rho}^{(\rho)}$. Here the threshold parameters were 
taken to be  $\Delta_s = \Delta_u  = 0.5$ GeV. 
Also in this case we find a good stability for a  
wide range of  $M^2$ values.  We have chosen the Borel mass to be   $M^2=3$ GeV$^2$. 
Having determined $M^2$,  we calculated the $Q^2$ dependence of the form factors. 
We present the results in Fig.~\ref{fig5}, where the squares correspond to the 
$g_{D^{*}D \rho}^{(D)}(Q^2)$ form factor in the  interval where the sum rule is valid. 
The triangles are the result of the sum rule for the $g_{D^{*}D \rho}^{(\rho)}(Q^2)$ 
form factor. 

\vspace{0.5cm}
In the case of an off-shell $ D $ meson, our numerical results can be 
fitted by the following monopolar parametrization (shown by the solid line in
Fig.~\ref{fig5}):
\begin{equation}
g_{D^{*}D \rho}^{(D)}(Q^2)= \frac{234.38} {Q^2 + 44.14} \;.
\label{monodoff}
\end{equation}
where the function $g_{D^{*}D \rho}^{(D)}(Q^2)$ has the units of GeV$^{-1}$, as we 
could anticipate from (\ref{lagran}). Following our previous works 
\cite{bclnn01,mnns02,mnns05,cdnn05,bcnn05}, 
we define the coupling constant as the value of the 
form factor at $Q^2= -m^2_{M}$, where $m_{M}$ is the mass of the off-shell meson. 
Therefore, using $Q^2=-m_{D}^2$ in Eq~(\ref{monodoff}), the resulting coupling 
constant is $g^{(D)}_{D^{*}D \rho } =   5.76 \, \GeV^{-1}$.

For an off-shell $\rho$ meson  our sum rule results  can  be 
fitted by an exponential parametrization, which is represented by the 
dot dashed line in Fig.~\ref{fig5}:  
\begin{equation}
g_{D^{*}D \rho }^{(\rho)}(Q^2)= 5.12 e^{-Q^2/4.33}\;.
\label{exprhooff}
\end{equation}
Using $Q^2=-m^2_{\rho}$ in Eq~(\ref{exprhooff}) we get 
$g^{(\rho)}_{D^{*}D \rho }= 5.89  \,  \GeV^{-1}$.

Looking at Fig. \ref{fig5}  we can observe that the $D$ off-shell form fator is much harder 
than the $\rho$ off-shell one. This agrees with the conclusions found in most of our 
previous works: the heavier is the off-shell meson, the harder is its form factor. Every 
extrapolation introduces some ambiguity in the final results, since we have the freedom to 
fit a set of points with different parametrizations. In our case {\it this freedom is 
strongly reduced because we require that   both parametrizations lead to the same coupling 
constant}. In Fig.   \ref{fig5} this requirement forces the two endpoints of 
(\ref{monodoff}) and (\ref{exprhooff}), which are taken at the squared masses of the 
corresponding  particles, to coincide, i.e., to have the same height in the figure. 

\begin{figure}
\vspace{1.0cm}
\centerline{\psfig{figure=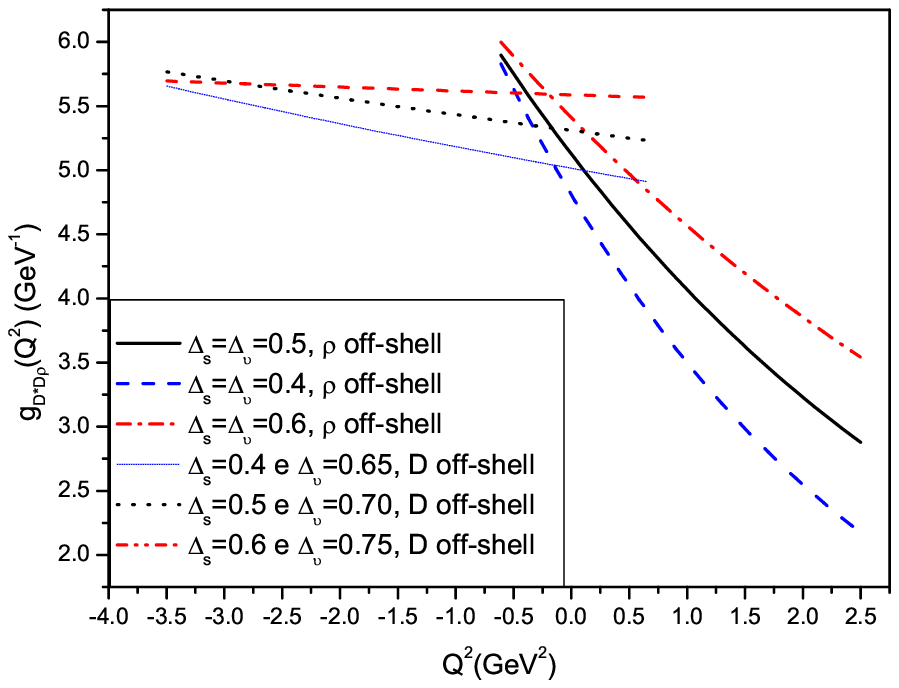,width=12cm}} 
\vspace{0.5cm}
\caption{Dependence of the form factor on the continuum thresholds. 
The solid curve  corresponds to 
$\Delta_{s,u} = 0.5\GeV$, the dashed one  to $\Delta_{s,u}= 0.6\GeV$
and the dotted curve to $\Delta_{s,u}= 0.4\GeV$.}
\label{fig6}
\end{figure}

In order to study the dependence of our results with the continuum
threshold, we vary $\Delta_{s,u}$ between 
$0.4 \GeV \le \Delta_{s,u} \le 0.6 \GeV$ in the  sum rule (\ref{ffrho}) and
$0.4 \GeV \le \Delta_{s} \le 0.6 \GeV$ and $0.65 \GeV \le \Delta_{u} \le 0.75 \GeV$
in the sum rule (\ref{ffd}). This variation produces new sets of curves which are shown in 
Fig.~\ref{fig6} and  gives us an uncertainty range in the resultig coupling constants. 
They are 
$g_{D^{*}D \rho }^{(D)}  = 5.70 \pm 0.06  \, \GeV^{-1} $ and 
$g_{D^{*}D \rho }^{(\rho)} = 5.90 \pm 0.08  \, \GeV^{-1}$.
We can see that the two cases considered here, off-shell $D $ or $\rho $, give 
compatible results for the coupling constant. Considering the uncertainties in the 
continuum thresholds and taking the average between the obtained values we have: 
\begin{equation}
g_{D^{*}D \rho }=  \Big (5.81 \pm 0.17  \Big ) \,  \GeV^{-1}
\label{finalcoupling}
\end{equation}

Our results were obtained for a concrete choice of currents, Eqs. (\ref{corho}),  
(\ref{coD}) and (\ref{coDs}), which represent charged states. Consequently the 
obtained couplings are for charged states and from them we can get the  generic 
coupling appearing in the Lagrangian (\ref{lagran}) through the relation:
\begin{equation}
g_{D^* D \rho }  =  \frac{g_{\rho^+ D^{0} D^{*+}}}{\sqrt{2}}
= \frac{g_{\rho^- D^{0} D^{*+}}}{\sqrt{2}} 
\end{equation}
Therefore the value of the coupling constant is:
$$ g_{D^{*}D \rho }= \Big ( 5.81  \pm 0.17 \Big ) /\sqrt{2}= 
\Big (4.13 \, \pm 0.08 \Big ) \, \GeV^{-1} $$

In Table II we compare this value with others obtained in previous works. 
For us the comparison between our results 
and those found in Ref. \cite{wang} and Ref. \cite{dai02} is especially meaningful, since 
both approaches use QCD sum rules, although in a different implementation. As it can be seen 
in Table II, these two works arrive at somewhat different values of the coupling constant, 
which are, within the errors, compatible with each other. We use the standard SVZ sum rules and 
the authors of \cite{wang,dai02} work with QCD Light Cone Sum Rules 
(LCSR). We use  the three-point function, whereas they use the two-point function 
with the $\rho$ as an external field. The advantage of using  the three-point function is that 
it allows us to treat  the $\rho$ meson as an off-shell particle and compute not only the 
coupling constant but also the form factor. Our results have non-perturbative corrections coming 
from condensates whereas in \cite{wang,dai02} the authors perform a twist expansion. 
In view of these differences it is reassuring to see that we obtain values of 
$g_{D^* D \rho}$ which are  compatible with each other.

\begin{table}[h]
\begin{tabular}{|c|c|c|c|c|} \hline
This work & LCSR \protect{\cite{dai02}}   &  LCSR  \protect{\cite{wang}}  & 
VDM  \protect{\cite{su}} & $SU(4)$ \protect{\cite{bcnn08}}
\\ \hline 
$4.11 \pm 0.08$ & $4.17 \pm 1.04$  &  $3.56 \pm 0.6$ & $2.82 \pm 0.1$ & $3.28 \pm 0.1 $ 
\\ \hline
\end{tabular}

\caption{$g_{D^* D \rho}$ in GeV$^{-1}$ obtained in previous  works.}
\label{tableres}
\end{table}

In Ref. \cite{su} the authors made an estimate of the 
${D^* D \rho}$ coupling constant applying the Vector Dominance Model (VDM) to the radiative 
decay $D^* \rightarrow D \gamma$ and using experimental information.  The obtained value is 
somewhat smaller than the others. We should take this estimate with caution, since it has been 
known since long ago \cite{huf} that the application of VDM to the charm sector is not always 
reliable. 

Another way to estimate unknown charm coupling constants is to connect them with  known 
couplings through $SU(4)$ relations. In the present case, we could use the relation:
\begin{equation}
g_{D^* D \rho }  =  \frac{g_{D^{*} D^{*} \rho }}{ m_{D^*}} 
= \frac{6.6 \pm 0.3}{2.01} 
=  \Big ( 3.28 \pm 0.15 \Big ) \,  \GeV^{-1}
\end{equation}
This number is smaller the QCDSR results. In our previous works \cite{mnns02,bcnn05} we 
found that, in QCDSR, the  $SU(4)$ relation 
$g_{J/\psi  D^* D^*} = g_{J/\psi D D}$ is satisfied.  
However, from \cite{bclnn01} and \cite{bcnn08} we observe that other  $SU(4)$ relations, 
such as $g_{\rho D^* D^*} = g_{\rho D D}$ and 
$g_{\rho D^* D^*} = \frac{\sqrt{6}}{4} g_{J/\psi  D^* D^*}$ are violated at the level of 
50 \%. This is not surprising since the mass difference starts to play an important role 
when we go from the  heavier vector mesons to $\rho$.

In conclusion, we have calculated the form factors of the $D^* D \rho$ vertex and also the 
coupling constant. We have used QCD sum rules to explore the properties of the three-point 
Green function of this vertex. The form factors $g_{D^{*}D \rho }^{(D)}(Q^2)$ 
(\ref{monodoff}) and  $g_{D^{*}D \rho }^{(\rho)}(Q^2)$  (\ref{exprhooff}) were obtained for 
the first time and, as mentioned in the introduction, they can be used in several 
phenomenological applications. The coupling constant  extracted from the form factors  is 
$g_{D^{*}D \rho } = 4.1  \pm 0.1$ GeV$^{-1}$ and it is in agreement with other QCDSR 
estimates.

\acknowledgments
This work has been supported by CNPq, CAPES and FAPESP.

\end{document}